\documentclass{sigchi}

\toappear{Permission to make digital or hard copies of all or part of this work for personal or classroom use is granted without fee provided that copies are not made or distributed for profit or commercial advantage and that copies bear this notice and the full citation on the first page. Copyrights for components of this work owned by others than ACM must be honored. Abstracting with credit is permitted. To copy otherwise, or republish, to post on servers or to redistribute to lists, requires prior specific permission and/or a fee. Request permissions from Permissions@acm.org.\\
{\emph{CSCW '16}}, February 27-March 02, 2016, San Francisco, CA, USA \\
\copyright~2016 ACM. ISBN 978-1-4503-3592-8/16/02...\$15.00 \\
DOI: http://dx.doi.org/10.1145/2818048.2819982}

\usepackage{balance}  
\usepackage{graphics} 
\usepackage{txfonts}
\usepackage{times}    
\usepackage[pdftex]{hyperref}
\usepackage{color}
\usepackage{textcomp}
\usepackage{booktabs}
\usepackage{todonotes}
\usepackage{comment}
\makeatletter
\def\url@leostyle{%
  \@ifundefined{selectfont}{\def\UrlFont{\sf}}{\def\UrlFont{\small\bf\ttfamily}}}
\makeatother
\def\pprw{8.5in}
\def\pprh{11in}

\definecolor{linkColor}{RGB}{6,125,233}
\hypersetup{%
  pdftitle={SIGCHI Conference Proceedings Format},
  pdfauthor={LaTeX},
  pdfkeywords={SIGCHI, proceedings, archival format},
  bookmarksnumbered,
  pdfstartview={FitH},
  colorlinks,
  citecolor=black,
  filecolor=black,
  linkcolor=black,
  urlcolor=linkColor,
  breaklinks=true,
}

\newcommand{\corrF}{Friends-Overlap}
\newcommand{\corrFshort}{FrOverlap}
\newcommand{\corrS}{Strangers-Overlap}

\newcommand{\pmep}{PME procedure}

\begin{document}

\title{Distinguishing between Personal Preferences and Social Influence in Online Activity Feeds}

\numberofauthors{2}
\author{%
  \alignauthor{Amit Sharma\\
    \affaddr{Dept. of Computer Science}\\
    \affaddr{Cornell University}\\
    \affaddr{Ithaca, NY 14853 USA}\\
    \email{asharma@cs.cornell.edu}}\\
  \alignauthor{Dan Cosley\\
    \affaddr{Information Science}\\
    \affaddr{Cornell University}\\
    \affaddr{Ithaca, NY 14853 USA}\\
    \email{drc44@cornell.edu}}\\
}

\maketitle

\begin{abstract}
Many online social networks thrive on automatic sharing of friends' activities to a user through activity feeds, which may influence the user's next actions. However, identifying such social influence is tricky because these activities are simultaneously impacted by influence and homophily.  We propose a statistical procedure that uses commonly available network and observational data about people's actions to estimate the extent of copy-influence---mimicking others' actions that appear in a feed.  We assume that non-friends don't influence users; thus, comparing how a user's activity correlates with friends versus non-friends who have similar preferences can help tease out the effect of copy-influence. 

Experiments on datasets from multiple social networks show that estimates that don't account for homophily overestimate copy-influence by varying, often large amounts. Further, copy-influence estimates fall below $1\%$ of total actions in all networks: most people, and almost all actions, are not affected by the feed. Our results question common perceptions around the extent of copy-influence in online social networks and suggest improvements to diffusion and recommendation models.

\end{abstract}



\section{Introduction}
In many social media, people consume and give feedback on items. 
For example, on Last.fm, people express their preferences for songs by \textit{listening} to or \textit{loving} them. 
These actions are shared to their friends or followers through an activity feed that aggregates the actions of a user's friends.

A fundamental question in these online social networks is how friends
influence people's preferences and decisions.  
Such influence operates through many mechanisms, including social proof, conformity, mimicry, and personal liking (see \cite{cialdini01} for a review) and 
can affect our consumption of information \cite{messing12,pariser11}, preferences for cultural items \cite{salganik06,sharma13}, adoption of new products \cite{aral09pnas}, behaviors around health \cite{centola13}, and even decisions to vote \cite{bond12}. Influence has also captured the popular imagination,  as a powerful force that makes content spread \textit{virally} through a social network \cite{gladwell06,christakis09}.  Measuring the extent of this influence, then, is an important problem in understanding diffusion in social networks. 

In general, observational data from social networks, both offline and online, do show correlation in friends' activities, or \textit{locality} in their preferences within a social network \cite{sharma-icwsm13}. Studies in  Twitter \cite{romero11}, Wikipedia \cite{crandall08}, Flickr \cite{cha09}, and Second Life \cite{bakshy09} consistently find that a user's probability of adopting an item increases with the number of friends who have done so before.

However, this locality may not be due to social influence because other social processes might also lead to the same correlation.  Notably, people are more likely to form social ties with people who are similar to them, via the homophily selection process \cite{mcpher01}. Thus, even in the absence of any social influence, underlying similarity between people may lead to observed similarities in their preferences on items that could lead to overestimates of social influence.

Untangling this knot is a problem that has vexed sociologists for decades.
In general, without either strong assumptions about influence mechanisms (e.g., \cite{steglich10,lewis12,christakis07}) or strong knowledge of latent variables that indicate homophily (e.g., \cite{aral09pnas}), a structural analysis of the causal graphical model \cite{pearl00} encoding influence and homophily shows that distinguishing them is impossible \cite{shalizi11}.
 
In this paper, we make principled assumptions about both the influence mechanism and indicators of homophily to develop a method for identifying influence that is well-suited for continuous activity data from online social networks. Since activity feeds are a primary interface element in most social networks that expose people to their friends' updates \cite{sun09,lerman10,bernstein13,bakshy11} and facilitate social learning \cite{burke09feedme,dabbish12}, we look specifically at how influence might flow through these feeds.
That is, we study whether
exposure to friends' activities through an aggregated feed makes a user more likely to copy or mimic them; we call this mechanism \textit{copy-influence} due to the feed.
For indicators of homophily, instead of using demographic or other latent attributes as in Aral et al. \cite{aral09pnas}, we directly use people's decisions on items to model their underlying preferences, and look at whether people copy friends' actions more than we would expect based on their underlying preferences.
That is, if a user sees that her friend on Last.fm loved a song, does it make her more likely to love it as well?  Or would she have loved it anyways? 

Our method is based on the observation that feeds rarely show activity of non-friends to a user. Thus, comparing how often a user copies actions from her friends' feed to how often she copies actions from a simulated feed constructed from actions of a matched set of non-friends with preferences similar to her friends provides a statistical procedure for estimating influence.
 
We first show that this procedure distinguishes copy-influence from preference similarity using semi-synthetic data based on a dataset we collected from Last.fm.  We then apply it to real activity data from Last.fm, Goodreads, Flixster, and Flickr.  Our results on this broad range of websites and item domains reveal three main takeaways about the role of copy-influence in online social networks. 

First, we find that naive estimates of copy-influence do overestimate it, often substantially, and that the degree of overestimation varies widely across datasets and kinds of actions. We argue that this variation is likely due to differences between sharing networks in characteristics such as the ease of consumption of items, the specific design of the feed, and its role in people's use of a site.  

Second, we find wide individual variation in susceptibility to copy-influence. Such variation extends past work on identifying differences in susceptibility \cite{cosley03,sharma13,aral12-susceptible}, and points to the potential of using individual-level susceptibility estimates for modeling diffusion and making more informative, persuasive recommendations using social information \cite{sharma13,kulkarni13}.

Third, we find that the extent of copy-influence is small.  Across all domains, less than 1\% of people's actions can be attributed to copy-influence.  Contrary to the popular narrative on influence, our results join other work in questioning the extent of influence or virality in these networks \cite{goel12,flaxman13}, at least as conveyed through copying others' observed behaviors in activity feeds.

This paper also contributes a way to estimate copy-influence in any network where network connections and timestamped activity data are visible, as well as a large Last.fm dataset.  More generally, we see this work as demonstrating the need for sharper definitions of influence that account for both personal preferences and specific mechanisms of influence.

\section{Related work}
Mimicry of others' actions has been demonstrated in controlled experiments. This can happen at a population level, as with Sagalnik et al.'s demonstration that making popularity visible causes different songs to become popular compared to the case when no such information is shown \cite{salganik06}.  It may be even stronger at the individual level; Sharma and Cosley showed that trusted friends' opinions have more influence on people's willingness to try unknown items than item popularity \cite{sharma13}.
Similar results have been shown in experiments on online social networks: a user is more likely to share a link on Facebook if it appears in her social feed \cite{bakshy12weakties} and to click on an ad if it is endorsed by her friends \cite{bakshy12ads}.  

Outside of controlled experiments around information consumption in real social networks---which may be infeasible, costly, or potentially unethical---identification of influence is not straightforward.
Naive measures such as simply counting the number of common actions between friends within a given time period likely overestimate influence \cite{bakshy11,ghosh10,tang13confluence}, as any observed data is simultaneously affected by both influence and homophily.  In addition, people may be \textit{externally exposed} to the same item through forces outside of the social network \cite{crandall08}.  For example, two friends may like the same item after watching an advertisement for it.  Such external exposure also creates confounds for estimating influence. 
 
In line with Shalizi and Thomas' argument for the need for assumptions about mechanism \cite{shalizi11}, work on estimating influence from observational data generally makes assumptions about the nature of influence and/or homophily, then uses statistical or computational procedures on those data to simulate contexts where those assumptions don't hold.

\subsection{Controlling for influence}
The first general strategy is to compare the observed data with an alternative world with no influence and attribute the differences to influence.  A core part of such an estimation procedure is to obtain, or create, data and network structures for the alternative world. 

Christakis et al. presented an edge-reversal test, where if person A has an edge to B but not vice-versa, then comparing B's influence on A with that of A on B would give us a measure of influence due to the directed edge \cite{christakis07}. This is based on the intuition that influence should flow only one way on directed edges; it also has the advantage of strongly addressing the effect of homophily.  This test was used to examine whether becoming obese is contagious and found that influence is significantly higher in the direction of the directed edge than the opposite direction.  In online contexts like Twitter with asymmetric information flows, this could fit well, but in networks with undirected edges, it is not so useful.

Another approach by Anagnostopoulos et al. randomizes the order of people's actions to remove causal links due to temporality between A's and B's actions \cite{anag08}. 
In the absence of influence, the expected probability of a user acting upon an item given that some number of their friends have already done so---called $k$-exposure \cite{romero11,crandall08}---should be the same in the observed data and the time-shuffled data.  If it is statistically different, influence is in play.
The method was used on a Flickr dataset and showed no significant difference between the two worlds.  Still, the authors do not rule out influence, giving examples from their dataset that demonstrate influence effects, and note that their method is unable to make individual-level influence estimates.
 
\subsection{Controlling for homophily}
Instead of removing influence, the other main strategy is to control for homophily.  For instance, La Fond and Neville use shuffling of social network edges to estimate influence given snapshots of people's network and activity data at two points in time \cite{lafond10}.  They first calculate the average overlap in activity between friends. To control for homophily, they subtract the overlap in friends' activity after randomizing the edges between people. To control for external exposure, they also subtract out the overlap among friends when both edges and actions are randomized.
Any difference that one finds then, must be due to influence.
Using data on Facebook groups joined by people at two timestamps a year apart, they found that the relative effect of influence varies: some groups exhibited a significant influence effect, while others exhibited a significant homophily effect. 

Another way to control for homophily is to directly account for its indicators.  If we are able to observe attributes such as demographics or other characteristics that affect people's preferences, then we can identify influence by controlling for these attributes.  Based on this intuition, Aral et al. estimated influence effects by comparing adoption rates for pairs of similar users where only one of them had a friend who had adopted the item \cite{aral09pnas}.  Using data from Yahoo on adoption of a web service and 46 attributes based on personal and network characteristics, they found that most adoption was independent of influence.
A fundamental problem, however, is that their method depends heavily on the careful choice and availability of attributes that predict similarity.

\section{Preliminaries: Basis for influence estimation}
Combining ideas from the above two lines of work, we propose a broadly applicable statistical procedure for estimating the extent of copy-influence in social networks. We control for homophily using activity data while limiting the time duration of exposure to friends' actions to closely model the mechanism of copy-influence through a feed interface.  In this, we make the following key assumptions.

\subsection{Preference as a proxy for homophily}
To control for homophily, we borrow from the recommender systems literature \cite{susurvey09} and assume that similarity metrics based on past activity are good indicators of underlying homophily. Past activity data implicitly captures factors such as demographics, prior external and social influence, and other hidden factors that determine people's preferences on items \cite{krulwich97,sandy13}; it can also be a direct indicator of their current personal preference. In principle, any two people with identical activity histories have the same probability of acting on an item in the future, minus any social influence.  

Using commonly available streams of activity data avoids cost and methodological issues around using panel data collected at fixed intervals
\cite{lewis12,steglich10,lafond10} and allows broader application of the matching technique from Aral et al. \cite{aral09pnas}, which required additional person-level and network attributes.

\subsection{Limited attention to a reverse chronological feed}
To specify the influence mechanism, we focus on copy-influence from feeds.  Feeds are not the only interface elements that might convey social influence: profile pages, collaborative filtering or popularity-based lists of items, social explanations of presented content, and out-of-system interaction may all convey information and perhaps influence from friends.  However, feeds are ubiquitous and commonly studied as conveyors of influence \cite{sun09,lerman10,bakshy11}; we argue that they are the dominant feature through which people see and learn from their  friends' behavior in social networks \cite{bernstein13,burke09feedme}, and thus a dominant feature through which influence from specific friends might flow.

\begin{figure}[t]
\center
\includegraphics[scale=0.9]{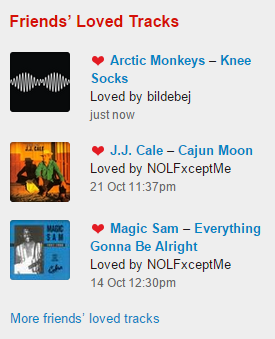}
\caption{A reverse chronological feed of songs loved by friends of a user on Last.fm. For each love action, the name of the song, its artist, and the friend who loved it are shown. This interface is shown as a widget on the home page for each logged-in user, along with a similar widget for recent songs listened to by friends. }~\label{fig:lastfm-interface}
\end{figure}

Further, we assume that the feeds show an unfiltered, reverse chronological aggregation of friends' activities. For many networks such as Last.fm, Goodreads, and Flickr, this closely resembles the feed seen by users. For example, Figure~\ref{fig:lastfm-interface} shows a reverse chronological feed of songs loved by the friends of a user on Last.fm.  Still, this is an approximation: feeds can contain advertisements or content from non-friends (e.g., sometimes Facebook presents actions from friends of friends), while algorithmic filtering can hide or reorder items shown in the feed.  We discuss ways to deal with such filtered feeds later.

\subsection{A simple model for users' interaction with feeds}
The above assumptions lead to a simple model for users' interaction with an activity feed.  Let us start by formally defining what we mean by a \textit{feed} within a social network. The term  \textit{friend} refers to anyone whose activities are shown to the user; these are typically social connections of an individual, such as Facebook friends or Twitter followees.
We define the feed to be the aggregated activity of all of a user $u$'s friends presented in reverse chronological order. 

We assume that at the time a user performs an action, she has seen some number $M$ of the friends' most recent actions in the feed.
In contrast to k-exposure models that implicitly assume people attend to all of their friends' actions \cite{anag08,backstrom06}, the parameter $M$ represents a user's attention budget for friends' actions and we argue this more realistically models copy-influence.  

Finally, we define \textit{copy-actions} as actions by a user on itmes that appear in the last $M$ items in her feed (e.g., loving one of the songs in Figure~\ref{fig:lastfm-interface} if $M>=3$).
A baseline measure of copy-influence would be the fraction of actions by a user that are copy-actions, as reported in some studies \cite{ghosh10,tang13confluence,bakshy11}. However, as argued earlier, that is likely to overestimate copy-influence because of underlying homophily and external exposure.

\section{Estimating copy-influence from feeds}
To address these possible confounds, let us now consider a hypothetical feed constructed from users who are not friends with the user. When a user's actions are copies of a feed based on non-friends, that is most likely due to following their own preferences or common external exposure.  Copy-influence is ruled out because within our model, a user does not see the non-friends' actions in her feed\footnote{Although direct copy-influence is ruled out, there may still be indirect, algorithmically mediated influence, such as through recommendations or targeted advertisements as described earlier.  Such influence may be important, but our focus here is on the extent of copy-influence.}. This intuition drives our procedure for estimating copy-influence: if we can find non-friends who are as similar to a user as that user's friends, we may use them to approximate the effects of homophily and some kinds of external exposure.  Thus, copy-influence can be estimated as the number of copy-actions from the friends' feed, over and above the observed overlap between a user's actions and the matched non-friends' feed.

To compute similarity between users, we consider their past actions on items as a representation of their preferences.
This allows us to use similarity metrics on past actions to find non-friends who are as similar in preferences to a user as her actual friends. Note that our selection of non-friends is similar to La Fond and Neville \cite{lafond10}, except that instead of choosing non-friends randomly, we use preference similarity to better control for homophily.  

In addition to controlling for homophily, using matched non-friends also controls for some kinds of external exposure. These would include mass  advertisements or widely popular items where any two users with the same personal preference would be expected to have an equal probability of acting on the item, regardless of the distance between them in the network\footnote{Like other methods, using matched non-friends cannot control for external exposure due to shared context (e.g., when two friends attend the same music concert); such overlap is much harder to distinguish from activity data alone.}. 

\subsection{Preference-based Matched Estimation (PME) Procedure}
The procedure proceeds in two phases: \textit{Matching} and \textit{Estimation}.  We divide the data into two parts at a fixed time $T$, performing the matching phase on data before $T$ and the estimation phase on data after $T$.  In the matching phase, for each user $u$ we generate a set of non-friends (``similar strangers") who at time $T$ are as close in preferences to the user as her friends.  In the estimation phase, we compute \corrF, the fraction of $u$'s actions that are copy-actions of a feed based on friends ($F$), and  \corrS, the fraction that are copy-actions of a feed based on the similar strangers ($S$).   The extent by which \corrF\ is greater than \corrS\ gives us a measure of copy-influence.

\begin{table*}[h]
 \center
 \begin{tabular}{lrr}
 \textbf{Feature} & \textbf{Listen} &\textbf{ Love} \\
Users with $\geq10$ actions & 312K & 434K  \\
Core users & 96K & 140K \\
Friends per user (mean; std. error; median) & (75; 0.7; 29) & (70; 0.5; 25)  \\
Actions & 656M & 136M\\
Actions per user (mean; std. error; median) & (2101; 5.5; 1192) & (313; 1.4; 118) \\
Songs & 23M & 13M \\
Actions per song & (28; 0.1; 2)  & (10; 0.03; 1) \\

 \end{tabular} 
 \caption{Descriptive statistics for the Last.fm dataset. On average, each user listened to 2101 songs during the 3 month period and loved a total of 313 songs during his lifetime.}
 \label{tab:desc-stats-lastfm}
\end{table*}

\subsubsection{Matching Phase} For each friend $f$ of a user $u$, we find a matching non-friend $s$ using activity data before time $T$, such that both of the following conditions are satisfied:
\begin{itemize}
 \item The similarity between the non-friend $s$ and $u$ is approximately equal to the similarity between $f$ and $u$. We compute similarity between two users by using the
Jaccard measure between their activity sets up to time $T$\footnote{We also tried cosine similarity, which gave comparable results; in principle, any reasonably efficient and unbiased similarity metric could be used.}.  Let $A_{0,T}^{(u)}$ denote the activity set of a user $u$ consisting of all the items she has acted on until time $T$. Then, we compute the similarity between two users $u$ and $v$ as:

$$ Sim(u,v) = J(A_{0,T}^{(u)}, A_{0,T}^{(v)}) = \frac{|A_{0,T}^{(u)} \bigcap A_{0,T}^{(v)}|}{ |A_{0,T}^{(u)} \bigcup A_{0,T}^{(v)}|}$$

 \item The number of actions by the non-friend $|A^{(s)}|$ is approximately equal to the number of actions by the friend $|A^{(f)}|$, up to time $T$. Assuming that the rate of activity stays the same before and after $T$, this condition ensures the non-friend and friend are expected to have an equal number of actions that will appear in $u$'s feed after $T$. 
 
\end{itemize}

To implement the matching phase, we sample a non-friend randomly without replacement and check whether it matches with an unmatched friend until there are no more unmatched friends left or no more non-friends to choose.  We allow matches to be approximately equal: $\epsilon_s$ is the allowed percentage difference in similarity and $\epsilon_a$ is the allowed percentage difference in the number of actions between non-friends and friends.

\subsubsection{Estimation Phase} For the data after time $T$, we compute the percentage of actions taken by $u$ that copied recent actions by either the set of friends ($F$) or similar strangers ($S$) that we computed in the matching phase. Because we assume that people have a finite amount of attention for their feed, we only consider the $M$ most recent actions by the set before each action of $u$.

More formally, let $A_{T,\infty}^{(u)}$ denote $u$'s activity set after time $T$, and $Feed_M^{(u,W)}$ denote the most recent $M$ actions taken by a set of users $W$ before $u$ acts on a given item.  We define the \textit{Overlap} between $u$ and the users in $W$ as:
$$ Overlap_{M}^{(u,W)} = \frac{ \sum_{a \in A_{T,\infty}^{(u)}}1\{ a \in Feed_M^{(u,W)}\}}{|A_{T,\infty}^{(u)}|}$$
where $1\{x\}$ represents the indicator function which is $1$ whenever $x$ is true and $0$ otherwise.

The difference in $Overlap$ between a user and her friends (\corrF) versus the similar strangers (\corrS) should give us an estimate of the copy-influence due to the friends' activity feed, over and above the homophily effects captured by similarity in preferences with the non-friends and any external influences that affect both friends and non-friends.
$$CopyInfluence_{u} = Overlap_{M}^{(u,F)} - Overlap_{M}^{(u,S)} $$
The mean of this per-user copy-influence estimate gives us an estimate of the extent of copy-influence in a social network.

\subsubsection{Interpretation}
An estimate close to $1$ implies that copy-influence is the dominant force driving people's actions, while an estimate close to zero implies that most of \corrF\ can be explained by underlying preference similarity. To measure uncertainty in the mean estimate, we compute the standard error using the bootstrap technique from resampling statistics \cite{efron86}.\footnote{Per-user mean copy-influence estimates are not independent because of how we  sample non-friends; thus it may not be appropriate to use the formula for standard error.}

The same procedure also provides us per-user estimates for susceptibility to copy-influence. The standard error for the per-user estimate can be calculated by repeating the estimation procedure with a new selection of non-friends each time.  Note that the copy-influence estimate can range between $-1$ and $1$.
When a user tends to adopt items that are popular outside their ego-network and few items from within the network, the estimate can be negative; we take this as an indicator that copy-influence is not present and thus should be considered as zero for that user.

\section{Validation on semi-synthetic data}
\label{sec:validate-lastfm}
To check the efficacy of the procedure in identifying copy-influence, we first run it on simulated worlds based on data from Last.fm where we fix the relative effects of copy-influence, personal preference, and external exposure. We use actual data in the matching phase and activity data simulated by these processes in the estimation phase, similar to the evaluation strategy used by La Fond and Neville \cite{lafond10}.

\subsection{Describing the Last.fm dataset}
Last.fm is a music service that records the songs that its users listen to on its website or supported desktop/mobile devices. Users can listen to, love, or ban songs. It also allows users to add other users as friends; both parties have to agree, so these links are undirected.
Each user sees the songs \textit{listened} to and explicitly \textit{loved} by her friends on the Last.fm homepage, allowing us to study the influence effects of both implicit and explicit actions on items. These actions are aggregated and presented in two reverse chronologically ordered widgets---one for friends' loves and one for listens---in the way our model of copy-influence assumes (see Figure~\ref{fig:lastfm-interface}).

We used Last.fm's API to collect listens, loves, and network data for users. Because the API does not provide timestamps for the creation of friendship links, we run the risk of incorrectly considering someone as a friend of a user before the actual link was made.  To reduce this problem, we randomly selected 1000 user ids of people who joined before 2010, with the thought that these older members of the system would tend to have more stable friend networks.
 
Starting with this seed set, we followed a weighted breadth-first search to obtain other users, adding friends to the search queue weighted by the number of already-found users they were friends with.  This weighting resulted in a reasonably well-connected component of the Last.fm social graph. The crawl was completed over the months of April-June 2014.\footnote{The full dataset can be accessed at \url{http://www.amitsharma.in/data/lastfm-social-activity.html}.}

In addition to the social network, we collected users' time\-stamped actions on items through February 2014.  For loves, we collected the user's entire history since they joined; for listening, which is much more frequent, we collected songs they had listened to starting in November 2013 to keep the dataset size reasonable.  The data crawl and subsequent analysis was judged by Cornell's institutional review board (IRB) as an exempt research protocol. 

For both listens and loves, we filtered out any users with less than 10 actions.
Table~\ref{tab:desc-stats-lastfm} shows some statistics for the data we collected. 
Although listening is a more frequent activity, the number of users with at least 10 listen actions is fewer than the corresponding number for the love action (Table~\ref{tab:desc-stats-lastfm}). This is because we collected listening data for only 3 months, so users who were inactive during the period will not appear in the listening dataset. 

Since we collected a sample of the network, some users are missing friends in the dataset.
We fetched the actual number of friends for each user and labeled users that have at least 75\%\footnote{We chose the threshold percentage $75\%$ as a tradeoff between having a representative sample of friends and being too restrictive for filtering. Results were similar at thresholds of 50, 90 and 100.}
 of their friends in the dataset as \textit{core users}. We apply the \pmep\ to only such core users, for whom we have a reasonable sample of their total friends.

\subsection{Generating semi-synthetic data} 
\label{validate}
The three processes are operationalized as follows:
\begin{itemize}
\item \textbf{Copy-influence}: Analogous to a reverse chronological feed, a user selects an item at random from a set containing the last $M$ items acted upon by her friends.

\item \textbf{Personal preference}: First, we select the $k$ most similar users to the current user by comparing the Jaccard similarity of their current activity sets. A user then selects an item at random from the last $M$ items acted upon by these $k$ most similar users.  We choose a low value of $k=10$ to find a narrow neighborhood of high-similarity users.

\item \textbf{External exposure}: We model such exposure by assuming that a user randomly selects the next item to act upon from the set of all items, weighted by their current popularity.

\end{itemize}

To generate synthetic data, we start from the state of the Last.fm loves dataset at time $T$, after which we replace the songs that users actually loved  with songs generated by either the copy-influence, personal preference, or external exposure process, while maintaining the original social network and timestamps of actions.
We also consider cases with a mixture of personal preference and copy-influence. For these cases, we fix the relative probability of selecting copy-influence over personal preference and each user decides which process to use for his next action based on this probability.

For our experiments, we set $M=10$ and selected time $T$ at three equally-spaced timestamps: 2014/01/01, 2013/10/01 and 2013/07/01. We choose values closer to our crawl date so the friend set that we use for estimation is close to the actual friend set. 
Both $\epsilon_s$ and $\epsilon_a$ were set to $0.1$, ensuring that non-friends are within $10\%$ of the corresponding similarity and activity for friends of a user. 
To get reliable estimates for similarity and copy-influence, we consider only those users who have at least 5 actions both before and after $T$. Finally, we generated data 100 times using each process and ran the \pmep\ for determining the extent of copy-influence in each.  Results shown are averaged across the 100 runs and across all three timestamps.

\begin{table}
\center
\begin{tabular}{p{8.4em}p{4em}p{4.20em}p{4em}}
\textbf{Process} & \textbf{\corrFshort} & \textbf{Copy-Inf.} & \textbf{Std. Err.} \\
Ext. Exposure (EE) & $0.0001$ & $2.4*10^{-5}$ & $8.1*10^{-6}$ \\
Personal Pref. (PP) & $0.04$ & $0.001$ & $0.0001$ \\ 
Copy-influence (CI) & $1.00$ & $0.95$ & $0.0004$ \\
CI-PP (50\%-50\%) & $0.53$ & $0.51$ & $0.0001$ \\
CI-PP (10\%-90\%) & $0.16$ & $0.12$ & $0.0001$ \\
CI-PP (1\%-99\%) & $0.05$ & $0.02$ & $0.0002$  \\ 

\end{tabular}
\caption{Sanity check on the proposed \pmep\ using loves on songs, showing \corrF, the copy-influence estimate, and the standard error of the estimate for each process. Each dataset simulates either external exposure, personal preference, copy-influence, or a mixture of personal preference and copy-influence. The \pmep\ correctly does not ascribe most of the copy-actions in the homophily and external exposure processes to copy-influence. For mixtures involving copy-influence, the procedure retrieves the true probability of copy-influence with a lower error than \corrF .}
\label{tab:sim-means}
\end{table}

\subsection{\pmep\ recovers simulated copy-influence}
As Table~\ref{tab:sim-means} shows, our copy-influence estimate is able to correctly rule out most correlated actions in the external exposure and personal preference processes, while it still shows a high copy-influence estimate
 for the copy-influence process. 

Likewise, when there is a mixture of personal preference and copy-influence, copy-influence estimates using the \pmep\ are closer to actual copy-influence than \corrF. It is still slightly higher than the  true extent of  copy-influence that we fixed for generating the data. The reason is that our matching may be inexact: for the personal preference process, a user should have the same overlap with the matched non-friends' feed as with the friends' feed, but in practice, our measure of preference similarity may not capture people's preferences completely.  These estimates are also noisy, especially for users with relatively little data.
Nevertheless, the observed error is much lower than that obtained with \corrF , showing that accounting for preference similarity is helpful in estimating copy-influence.    

\section{Extent of copy-influence on Last.fm}
We now apply the PME procedure to the actual activity of users on Last.fm for both love and listen actions.
We use the same parameters as we did before, setting $M=10$ and $\epsilon_s=\epsilon_a=0.1$. Since we have listen data only for three months, we set $T$ differently for listen and love actions.  For love actions, we set $T=2013/07/01$ as before, while for listen actions, we set $T=2014/01/01$. 

A user may listen to the same song more than once, raising the question of how to treat repeated activity.  One option would be to only look at the first time a user heard a song, on the assumption that copy-influence plays a minimal role in re-experiencing the song versus a user's own preferences about the song after listening to it.  On the other hand, a user might be influenced to re-listen to a song they like by seeing it in their feed; in this case, we would want to measure copy-influence on all actions. For our copy-influence estimates, we consider all actions taken by users, including re-listens.

\begin{table}
\center
\begin{tabular}{llll}
\textbf{Action on song} & \textbf{\corrFshort} & \textbf{Copy-Inf.} & \textbf{Std. Err.} \\
Listen-Listen & $0.004$ & $0.001$ & $2.5*10^{-5}$ \\
Love-Love & $0.015$ & $0.003$ & $7.8*10^{-5}$ \\
Love-Listen & $0.004$ & $0.0007$ & $4.9*10^{-5}$ \\
\end{tabular}
\caption{A comparison of copy-actions from the friends' feed (\corrF) with the copy-influence estimate computed using the \pmep\ on Last.fm.  We look at the influence of friends' listen actions on a user's listen actions, love actions on love, and friends' love actions on a user's listen actions. For all three actions, \corrF\ overestimates copy-influence.}
\label{tab:actual-means}
\end{table}

\subsection{\corrF\ overestimates copy-influence}
Our first major observation is that for both listens and loves, \corrF\ overestimates copy-influence. Table~\ref{tab:actual-means} shows the mean effect of copy-influence, along with \corrF, for listening to or loving songs. On average, only $0.3\%$ of a user's actions can be attributed to influence for loves, and fewer still for listens---and in both cases, the copy-influence estimate is less than one-fourth of the naive \corrF .\footnote{In practice, to make the feed manageable, Last.fm's listen feed deviates from our reverse chronological assumption and instead shows one most recent song per friend. When we modify our feed model to include only the most recently listened song by each
friend in the feed, we find that the copy-influence
increases (from 0.001 to 0.0018), but is still less than half of \corrF .} 

To test whether explicit endorsement of a song by friends influences a user's implicit listening activity, we next compute copy-influence on listen actions of a user due to love actions in her feed.  As for the previous two scenarios, we find that \corrF\ for listens overestimates copy-influence and the actual extent of that influence is below $0.1\%$.

\corrF\ and copy-influence for exposure to loves is higher than those for listens, indicating first, that a higher fraction of love actions are common among friends than listens, and second, a higher fraction of love actions are also copied from friends than listens. A possible reason could be that loves by other users might be considered as stronger endorsements than the implicit listens, and thus, have higher copy-influence.  Further, there might be less user choice in which songs to listen to versus loves, for instance, when a playlist or recommendation algorithm is choosing which songs are played.  We will come back to how intentionality, along with algorithm and interface design, are important for modeling influence in the discussion.

\subsection{Most users show no influence effects}
In addition to estimating network-level copy-influence, it could be useful to estimate copy-influence on individual users. One way to interpret individual estimates is that we are measuring the \textit{susceptibility} to copy-influence for an individual \cite{aral12-susceptible}. Such an estimate can be used in diffusion models to set personalized transmission probabilities or thresholds.

\begin{figure}[t]
	\center
	\includegraphics[scale=0.7]{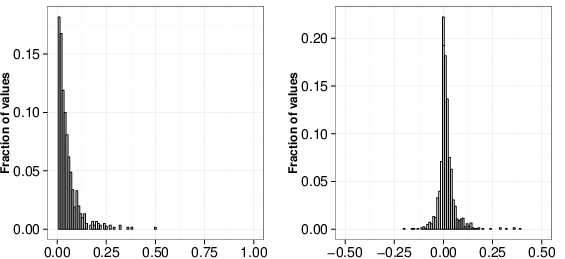}
	\caption{Per-user variation of \corrF\ (left panel) and copy-influence (right panel) for loves on songs among users with non-zero \corrF. Even among these users, copy-influence is zero or negative for a third of the users.
	}
\label{fig:inf-inbound-distr}
\end{figure}

Looking at per-user copy-influence for the \textit{love} action on songs shows a wide variation in the effect of copy-influence among users, as shown in Figure~\ref{fig:inf-inbound-distr}.  Note that strikingly, \corrF\ is zero for a majority of users.  The high number of zeros makes visual inspection of the distributions difficult, so Figure~\ref{fig:inf-inbound-distr} includes only users with non-zero \corrF .

Among users with non-zero \corrF, about a third have their copy-influence estimate less than or equal to zero, which indicates that all of their \corrF\ can be explained by preference similarity alone; overall, the copy-influence estimate is zero or negative for more than $75\%$ of core users. This helps to explain the low overall effect we see for copy-influence and provides additional empirical evidence of a wide range of susceptibility among users \cite{anag08,aral12-susceptible}.

\subsection{Less active users are more susceptible}
\begin{figure}[t]
	\center
	\includegraphics[scale=0.4]{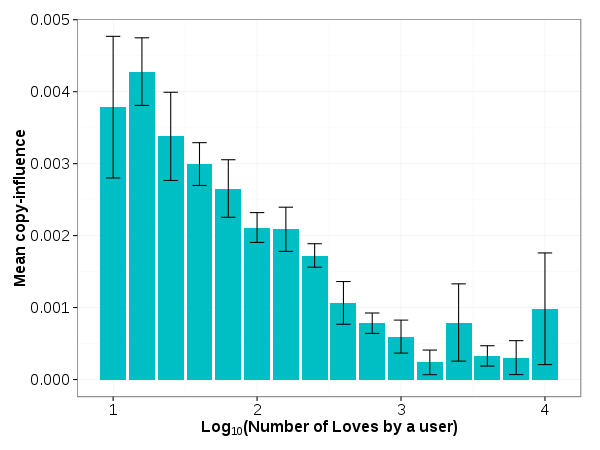}
	\caption{Variation of copy-influence with activity rates of Last.fm users during the Estimation phase. We filtered out bins with less than 5 users towards the right end of the plot. The effect of copy-influence decreases as the activity level of a user increases. Error bars convey standard error of the mean estimates.
	}
\label{fig:activity-susc}
\end{figure}

We also wondered whether users' activity levels affect how susceptible they are to copy-influence.  To do this, we compare copy-influence estimates for users with different numbers of loves on Last.fm during the Estimation phase. Figure~\ref{fig:activity-susc} shows the variation of copy-influence on songs with activity (love) rate of users. The extent of copy-influence is higher for users with fewer loves, who are also the majority of users in the dataset; as the activity level increases, copy-influence estimates decrease. 

\begin{figure}[t]
	\center
	\includegraphics[scale=0.4]{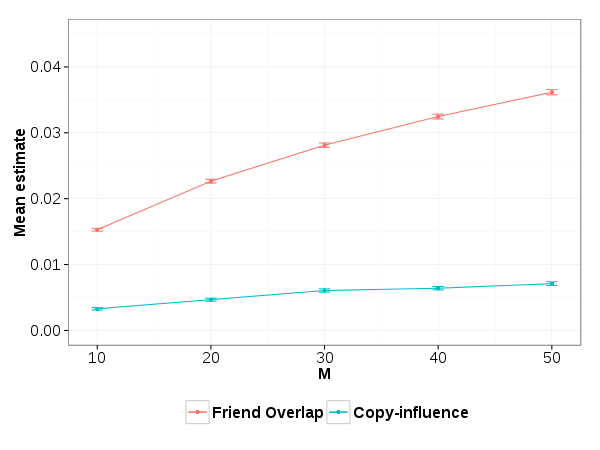}
	\caption{Variation of copy-influence and \corrF\ estimates with change in $M$. Both estimates increase; however, the ratio between the two remains approximately equal.
	}
\label{fig:inf-m-vary}
\end{figure}

\subsection{Robustness with change in parameters}
The parameters we use in the influence test are $T$, $M$ and the two error thresholds, $\epsilon_s$ and $\epsilon_a$. To check whether our results are robust to changes in parameters, we tried a range of values for each. While we do find changes in the influence estimate (increasing $\epsilon_s$ and $\epsilon_a$, for example, decreases the extent to which we can account for homophily), the high-level finding that copy-influence could be ruled out for most copy-actions stays consistent. 

Figure~\ref{fig:inf-m-vary} shows that both \corrF\ and estimated copy-influence increase with $M$.  The fact that both increase is not surprising, since as $M$ increases, each of a user's actions might match with a larger pool of actions from either the friends or non-friends feed.  However, the ratio between the two remains the same.

\begin{table*}[h]
 \center
 \begin{tabular}{l@{\hskip 0.01in}c@{\hskip 0.05in}c@{\hskip 0.05in}c}
 \textbf{Feature} & \textbf{Goodreads} &\textbf{ Flixster} & \textbf{Flickr} \\
Users with $\geq10$ actions & 252K  & 50.0K & 183K  \\
Users with friendship data & 252K & 48.8K& 175K\\
Friends per user & (29; 0.2; 14)  & (13; 0.1; 5) & (74; 0.5;17)   \\
Actions & 28M & 7.9M & 33M\\
Actions per user & (112; 0.35; 63) & (158; 1.6; 44) & (183; 1.5; 44)\\
Items & 1.3M & 48.4K & 10.9M\\
Actions per item & (21; 0.4; 2) & (163; 4.0; 5) & (3; 0.003; 1)\\

 \end{tabular} 
 \caption{Descriptive statistics for datasets from Goodreads, Flixster, and Flickr. Flixster has the lowest number of items and consequently, more actions per item (163) than Goodreads (21) and Flickr (3).  The average number of actions per user is above 100 for all three datasets.}
 \label{tab:desc-others}
 \end{table*}
 
\section{Copy-influence on other social networks}
\label{other-datasets}
We have focused so far on Last.fm as a running example for explaining and exploring the \pmep . We now apply it to compute copy-influence estimates on other social networks and see how our results generalize to different item domains, feed interfaces, and system designs.  Table~\ref{tab:desc-others} presents aggregate data for three existing datasets we use from Goodreads \cite{huang12}, Flixster \cite{jamali10flixsterdata}, and Flickr \cite{cha09}. 
These datasets cover a diverse set of item domains: Goodreads is a social network for books, Flixster for movies, and Flickr for photos.  More details about each dataset can be found in the papers that introduced them \cite{huang12,jamali10flixsterdata,cha09}.

\subsection{Describing the datasets}
As on Last.fm, users on each of these websites can act upon items and form undirected connections with other users. Goodreads and Flixster allow users to \textit{rate} books and movies respectively. Flickr allows users to \textit{favorite} photos that other users post on the website.  
 In addition, each social network  has a feed interface that shows friends' rating or favoriting activities, aggregated and presented in a loosely reverse chronologically order in the way our model of copy-influence assumes\footnote{Flixster moved away from being a social network for movies after 2010, but the current dataset was collected before the change and satisfies our assumptions.}. 

The number of items available to act on among the three datasets varies widely. On average, an item is acted on 3 times on Flickr, 21 times on Goodreads, and 163 times on Flixster. For comparison, the same average was 10 loves per song on Last.fm (in this section, we report only loves on Last.fm because they are more similar to rating or favoriting actions in these websites).  In contrast to per-item activity, per-user activity is similar across all three datasets, with users acting on over 100 items on average.  

Since on Goodreads and Flixster a friend's rating is shown in a user's feed irrespective of whether it was high or low, we consider all ratings by users for our analysis\footnote{To better account for preference similarity, we also tried a variation where we filtered out any rating below 3 or 4 on a scale of 0.5-5. The pattern of results is the same.}.
Since we do not know the real number of friends for each user in these datasets, we consider any user with non-zero friends as a core user.  All friend relations are still without timestamps, so we assume a static social network and set $T$ for each dataset so that only $10\%$ of the actions are after time $T$. For consistency, we use the same values for all other parameters as for Last.fm.

\begin{figure}[t]
	\center
	\includegraphics[scale=0.45]{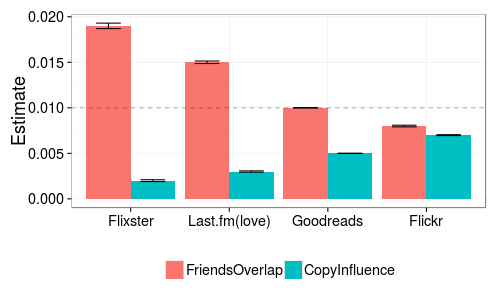}
	\caption{\corrF\ and copy-influence estimates on different social networks. While both \corrF\ and the ratio of \corrF\ to copy-influence varies, the copy-influence estimate for all websites is below 1\% of users' actions on items.}
	\label{fig:influence-low}
\end{figure}

\subsection{Vast majority of actions are due to personal preference}
Figure~\ref{fig:influence-low} shows the \corrF\ and copy-influence estimates for all four datasets, along with standard error. 
We find varying estimates for \corrF\  on these datasets, from $0.019$ on Flixster to $0.008$ on Flickr. The amount of \corrF\ explained by preference similarity also varies widely; the copy-influence estimate is less than 15\% of \corrF\ for Flixster and over 85\% of \corrF\ for Flickr. Such differences are plausible, and in fact, expected: except for the fact that they are all online social networks, these websites differ from each other in many characteristics: their item domains, types of actions, popularity of items, user interfaces, and use cases. 

However, after controlling for preference similarity between friends,  we find that the copy-influence estimates for all four domains fall into a range from 0.002--0.007.  
Interpreting the copy-influence estimate as the average fraction of user actions due to copy-influence, these results imply that less than $1\%$ of the total user actions in these sites can be attributed to copy-influence from the feed.
 
\section{Discussion}
\label{sec:discussion}
We present PME, a broadly applicable estimation procedure for separating social influence from preference behavior in online social networks. Unlike past work on identifying influence \cite{anag08,aral09pnas,christakis07}, our procedure does not need any additional person-level attributes, relies on data that is often publicly available, does not depend on the directionality of edges, and provides both overall and person-level estimates of not just the presence but the amount of copy-influence.  Applying this procedure to data from a broad range of websites shows that simple correlational estimates between friends greatly overestimate the extent of copy-influence and that accounting for preference similarity can lead to more accurate estimates.  Our results add depth to studies that found that a majority of actions common between friends can be attributed to homophily instead of influence \cite{aral09pnas, lewis12}, both showing that it occurs in a number of contexts and that the extent of overestimation varies widely with both users and systems.

We begin by placing our results in context with scholarly work on diffusion and influence in online social networks, questioning whether the effects of copy-influence are overrated. We then discuss plausible reasons for differences in the  overestimation of actual copy-influence that we saw for different social networks.
Finally, we revisit assumptions of the \pmep\ that should be kept in mind before using it in other settings.

\subsection{Copy-influence is overrated (?)}
Our results indicate a subdued picture of the role of copy-influence in online sharing networks.  In all four datasets, more than 99\% of people's actions on items can be explained without copy-influence, and most users are not at all influenced by the activity feed.  Even without controlling for homophily, \corrF---an upper bound on the actual copy-influence---is low: less than 2\% of users' actions are copy-actions (Figure~\ref{fig:influence-low}).  

These results are at odds with both popular perception and much scholarly work on social influence.
The idea of a particular item or opinion starting from a few sources and spreading \textit{virally} to a large component of a social network is riveting and often highlighted in popular discourse.  Further, social influence  on online networks has been shown to be present for a number of activities, as we pointed out earlier.  Thus, we are not arguing that influence is not important.

However, we do think that in many contexts it is likely overstated.  In particular, our results call attention to the tiny role of copy-influence in affecting people's routine activity around items.  Most discussion of influence focuses on the unusual: events where people do adopt their friends' behaviors and items spread widely.  
However, the mundane reality is that most people's actions are driven by their own preferences. Recent empirical work on the spread of items through social media provide further support for this claim. Flaxman et al. show that only about 2\% of online news consumption is referred through social media \cite{flaxman13}.  This is essentially an upper bound on the fraction of news articles read due to influence from others in a social network.  This study, like our work, is able to put the effect of social media on news consumption in perspective by comparing it against the total number of news articles visited by users and not just focusing on the ones shared within social networks.  Likewise, the vast majority of tweets never breach the sender's ego network \cite{goel12}.

Still, a small effect today could turn into a big one tomorrow:
$1\%$ of all actions seems tiny,  but $1$ influenced action out of $100$ may aggregate over time to be a significant effect on people's decisions.  This might be especially true if those influences happen earlier in a user's lifetime, which Figure~\ref{fig:activity-susc} suggests is true, since lower-activity users appear to have higher copy-influence estimates.  Studying the temporal variation in copy-influence estimates and evolution of people's personal preferences over time will be interesting future work.

\subsection{Factors that affect the extent of copy-influence}
We also observe variations in the degree of over-estimation for copy-influence between different actions (listen and love on Last.fm) and between different websites, and it will be useful to understand the reasons for these differences.  We see two factors as playing a prominent role in affecting the potential influence of friends' actions: characteristics of the item space and the design of the feed.  Our goal here is not to present an exhaustive account of reasons why copy-influence might differ between domains and individuals (in particular, network factors such as tie strength and knowledge of friends also affect people's copying decisions \cite{sharma13,aral14,bakshy12weakties,bond12}), but instead to call attention to the need for nuance in thinking about the mechanisms by which influence flows.

\subsubsection{Characteristics of the item space}
One set of factors that might affect the extent of copy-influence has to do with properties of the items and domain. 
\begin{itemize}
\item
 The distribution of attention to items likely matters.  For example, photos on Flickr, being user-generated items, are numerous compared to relatively mass-consumed items such as movies, music, and books. This leads to the low mean popularity of $3$ for a photo that we see in Table~\ref{tab:desc-others}. Consequently, the likelihood that people are exposed a particular photo through their friends rather than other mechanisms is higher. There could be a similar effect for explicit loves on Last.fm, which are rarer than listens and signal stronger interest from others.
\item 
Further, items like books, movies, and songs have a well-defined existence outside of the social networks we study that might lead to more external exposure.  Photos, on the other hand, often exist only on Flickr and thus are hard to discover from outside the system itself. 
\item
Finally, photos, like songs, are quick consumption items and thus are more amenable to mimicry on exposure from friends' feeds than a book or a movie.  A user may favorite a photo right after viewing it in her feed or love a song after listening to it for a few minutes. For domains like books or movies which take hours to consume, we would expect less such spontaneous mimicry. 
\end{itemize}
These characteristics of the item space can affect both the amount of overlap in activity between friends and the actual copy-influence flowing between them. Table~\ref{tab:compare-websites} summarizes these factors for the four websites we studied.  
The column for Flixster in the table suggests why movie ratings on Flixster have low copy-influence: the dataset does not possess any of the three factors that promote the likelihood of copy-influence. On the other hand, Flickr, where preference similarity could rule out only a small fraction of \corrF, satisfies all these criteria. Last.fm and Goodreads lie the middle. 

\begin{table}
\begin{tabular} {l@{\hskip 0.1in}c@{\hskip 0.05in}c@{\hskip 0.05in}c@{\hskip 0.05in}c}
\textbf{Item properties}  & \textbf{Flickr} & \textbf{Goodreads} & \textbf{Last.fm}   & \textbf{Flixster} \\
 No external existence & Yes & No & No & No \\
 Sparse actions/item & Yes & Yes & Yes & No \\ 
 Quick consumption  & Yes & No & Yes & No \\
\end{tabular}
\caption{Item properties for the four datasets. Flickr, which has the highest copy-influence estimate, has each of the properties that we argue promote copy-influence. Meanwhile, Flixster does not satisfy any of these properties and has the lowest copy-influence estimate despite having the highest \corrF\ among the four websites.}
\label{tab:compare-websites}
\end{table}

\subsubsection{Properties of the feed}
These websites also differ in how they show friends' activities to a user. Since users are typically exposed to their friends' activity through the feed, both the design of the feed interface and ranking algorithms that perturb the feed likely impact the extent of copy-influence in a network. 

Figure~\ref{fig:lastfm-interface} shows the feed for love actions on Last.fm; in the grand scheme of the interface, however, it is a small widget.  Moreover, the interface is often in the background while people listen to music and attend to other tasks, likely reducing the feed's influence relative to other sites.
The interface for Flickr lies on the other side of the spectrum: most of the available space is devoted to a feed of others' photos, so using the interface is a focal activity that concentrates attention on the photos shown in the feed.  

Feeds might also make acting on an item possible without having to consume it.  For instance, a user might not read a book or watch a movie right away, but if the interface lets her put it in a queue, this might reasonably be considered copy-influence on the \textit{queuing} action.  Note that in this case, the design of the feed and the actions afforded to users compensates for the item property of slow consumption. 

This discussion underscores the importance of precise definitions and explicit mechanisms for interpreting the influence estimates we obtain. System design and characteristics of the item space both impact people's exposure to and ability to act on items from their friends, thus impacting what we \textit{measure} as copy-influence in these social networks. 
Even though the \pmep\ is broadly applicable to social networks---requiring only social network edges and timestamped preference data---it is important to interpret the resultant influence estimates with respect to the design and context of the specific network. 

\subsection{Assumptions and Generalizability}
We now return to the assumptions behind the \pmep\ and discuss how it may be adapted to estimate copy-influence from feeds in other social networks.  In particular, we assume limited attention to a reverse chronological feed, focus only on influence conveyed by the feed, consider similarity in actions as a proxy for homophily, assume a static network, and equally weight activites by all of a user's friends.

\subsubsection{Limited attention to a reverse chronological feed}
Although the feed interfaces of the four websites differ somewhat, at heart they are based on a reverse chronological feed, leading to our assumption of such a feed, cut off by a threshold $M$ number of actions that users attend to. 
 When applying the influence estimation procedure to a new website, it is important to think about whether the specific feed ranking and interface of a website are amenable to such a formulation.  
 
For example, the reverse chronological assumption is less appropriate for networks with opaque feed ranking algorithms such as Facebook. In such cases it would be important to capture data about which items are actually shown to a user or use knowledge about the algorithm to approximate the real feed from the timestamped activity data. In addition, future work on estimating personalized values of $M$ for each user might lead to more fine-grained estimates of susceptibility and copy-influence.

\subsubsection{Copy-influence, not all influence}
Our definition of copy-influence means that we are only capturing kinds of influence that directly lead to the consumption of an item shortly after being exposed to it by a friend.  In the context of narratives around information diffusion and virality, it is a reasonable choice: to propagate through the network, users must continue to interact with it in ways their friends will see (and copy).  Further, we argue that this definition, by providing a concrete mechanism and measure for specific aspects of influence, is useful for carving the broad notion of ``influence'' into pieces that support better modeling, theorizing, and design.   

But it does mean that we are not talking about all mechanisms or methods of influence.  Focusing on the feed means we are not studying influence that flows through other interface features; focusing on copying of specific actions means that we are not looking at indirect or cumulative notions of influence (e.g., k-exposure models).  Thus, it is important to remember that although the \pmep\ does its job in identifying copy-influence and shows the importance of personal preferences, it does not capture all aspects of influence in a system.

\subsubsection{Preference and matching as a proxy for homophily}
We use preference similarity as a proxy for modeling underlying homophily between people.  To do this, one needs to have sufficient data to make reasonable preference models; this appears to be the case in many online social networks in which users generate large volumes of activity.  Since the effectiveness of the PME procedure hinges on the ability to match non-friends to friends, larger networks, higher activity levels, and context-appropriate metrics of similarity are likely to improve the reliability of copy-influence estimates.

We also chose to account for personal preference by matching friends with non-friends.  A natural alternative would be to directly compute a user's affinity for an item (e.g., using a recommendation algorithm \cite{susurvey09}) and use that to control for a user's own preference.  
However, the drawback is that the interpretation of influence estimates would depend strongly on the quality of the recommender algorithm as a proxy for personal preference, while such a recommender would not be able to account for external influences that might be evident in other users' activities. 

\subsubsection{A reasonably static network}
The datasets we studied didn't have timestamps for tie formation, so we assumed a static social network.  When timestamps for edge formation are available, we can make a simple modification in the procedure to consider only people who became friends before time $T$ when computing the set of similar strangers, then considering the current friends of a user when computing \corrF.  

Another limitation of our estimation procedure is that by considering all actions in the Matching phase (before time $T$) as proxy for personal preference, we miss out on the effects of copy-influence in those actions. If copy-influence is higher early on, we might underestimate it, as to the \pmep\ the effects of early copy-influence would look like personal preference at time $T$. 
Models that relax the time $T$ assumption and compute both preferences and friend sets across the history of the dataset are computationally expensive, but possible---and would be interesting for bringing simulation-based results around contagion and network change \cite{dodds05,doreian13} toward real datasets.  Further, better understanding how users' susceptibility changes as a function of their time in the system and their networks is an interesting question in its own right.

\subsubsection{All friends being equal}
Finally, by comparing all friends of a user in aggregate with their matched non-friends, we do not consider possible differences in how much copy-influence a particualr friend might wield.  Past studies show that people perceive actions by different friends differently based on their relationship with them \cite{aral14,cosley03,kulkarni13,sharma13}.
In our current datasets, we did not have any principled ways to estimate tie strength. Future work would include accounting for these tie-specific variations in influence.

\section{Conclusion}
We presented a statistical procedure for separating copy-influence from behavior rooted in personal preferences in observational data. The procedure requires only activity and network data for users, thus making it well-suited for many online social networks. At least for the websites we studied, applying the influence-estimation procedure shows that the vast majority of people's actions on items in social networks can be explained without invoking copy-influence.

More generally, our results suggest the importance of having clear definitions of influence and well-defined behavior and process models.  Rather than a grand unifying theory of all influence, we believe that focusing on specific processes are more likely to be useful in producing deeper understanding and models of influence.  Our results further suggest that past user activity can be a valuable resource for these models, requiring minimal additional data about users while giving valuable information about their underlying preferences.  

Accurate estimates of influence, in turn, promise to improve both models of behavior and design for online social networks.  Individual-level estimates can lead to better models of susceptibility for diffusion models and  better prediction of next actions of users for personalization tasks.  They would also enable evaluation of changes to feeds and interfaces and be practically useful for improving user experiences on online social networks.

\section{Acknowledgments}
This work was supported by the National Science Foundation under grants IIS 0910664 and IIS 1422484, and by a grant from Google for computational resources. We thank Chenhao Tan for helpful feedback on an early draft of the paper.

\balance{}

\bibliographystyle{SIGCHI-Reference-Format}

\end{document}